\documentclass[twocolumn,english]{revtex4-1}
\usepackage[T1]{fontenc}
\usepackage[latin9]{inputenc}
\usepackage{textcomp}
\usepackage{amsmath}
\usepackage{amssymb}

\makeatletter
\usepackage{babel}

\usepackage{babel}

\usepackage{babel}

\usepackage{babel}

\usepackage{babel}

\makeatother

\usepackage{babel}
\begin{document}
\global\long\def\abs#1{\left|#1\right|}
 \global\long\def\ket#1{\left|#1\right\rangle }
 \global\long\def\bra#1{\left\langle #1\right|}
 \global\long\def\half{\frac{1}{2}}
 \global\long\def\partder#1#2{\frac{\partial#1}{\partial#2}}
 \global\long\def\comm#1#2{\left[#1,#2\right]}
 \global\long\def\vp{\vec{p}}
 \global\long\def\vpp{\vec{p}'}
 \global\long\def\dt#1{\delta^{(3)}(#1)}
 \global\long\def\Tr#1{\textrm{Tr}\left\{  #1\right\}  }
 \global\long\def\Real#1{\mathrm{Re}\left\{  #1 \right\}  }
 \global\long\def\braket#1{\langle#1\rangle}
 \global\long\def\escp#1#2{\left\langle #1|#2\right\rangle }
 \global\long\def\elmma#1#2#3{\langle#1\mid#2\mid#3\rangle}
 \global\long\def\ketbra#1#2{|#1\rangle\langle#2|}

\title{From curved spacetime to spacetime-dependent local unitaries\\
 over the honeycomb and triangular Quantum Walks}

\author{Pablo Arrighi}
\email{pablo.arrighi@univ-amu.fr}

\affiliation{Aix-Marseille Univ, Université de Toulon, CNRS, LIS, Marseille, France
and IXXI, Lyon, France}

\author{Giuseppe Di Molfetta}
\email{giuseppe.dimolfetta@lis-lab.fr}

\affiliation{Aix-Marseille Univ, Université de Toulon, CNRS, LIS, Marseille, France
and Departamento de F{í}sica Te{ó}rica and IFIC, Universidad de
Valencia-CSIC, Dr. Moliner 50, 46100-Burjassot, Spain}

\author{Iván Márquez-Martín}
\email{ivan.marquez@uv.es}

\affiliation{Aix-Marseille Univ, Université de Toulon, CNRS, LIS, Marseille, France
and Departamento de F{í}sica Te{ó}rica and IFIC, Universidad de
Valencia-CSIC, Dr. Moliner 50, 46100-Burjassot, Spain}

\author{A. Pérez}
\email{armando.perez@uv.es}

\affiliation{Departamento de F{í}sica Te{ó}rica and IFIC, Universidad de Valencia-CSIC,Dr. Moliner
50, 46100-Burjassot, Spain}
\begin{abstract}
A discrete-time Quantum Walk (QW) is an operator driving the evolution
of a single particle on the lattice, through local unitaries. Some
QW admit, as their continuum limit, a well-known equation of Physics.
In \cite{PhysRevA.97.062111} the QW is over the honeycomb and triangular
lattices, and simulates the Dirac equation. We apply a spacetime coordinate
transformation upon the lattice of this QW, and show that it is equivalent
to introducing spacetime-dependent local unitaries \textemdash whilst
keeping the lattice fixed. By exploiting this duality between changes
in geometry, and changes in local unitaries, we show that the spacetime-dependent
QW simulates the Dirac equation in $(2+1)$\textendash dimensional
curved spacetime. Interestingly, the duality crucially relies on the
non linear-independence of the three preferred directions of the honeycomb
and triangular lattices: The same construction would fail for the
square lattice. At the practical level, this result opens the possibility
to simulate field theories on curved manifolds, via the quantum walk
on different kinds of lattices. 
\end{abstract}

\date{\today}

\maketitle

\section{Introduction}

{\em Quantum walks.} QWs are dynamics having the following characteristics:
\emph{(i)} the state space is restricted to the one particle sector
(a.k.a. one `walker'); \emph{(ii)} spacetime is discrete; \emph{(iii)}
the evolution is unitary; \emph{(iv)} the evolution is homogeneous,
that is translation-invariant and time-independent, and \emph{(v)}
causal (a.k.a. `non-signalling'), meaning that information propagates
at a strictly bounded speed. Their study is blossoming, for two parallel
reasons.\\
 One reason is that a whole series of novel Quantum Computing algorithms,
for the future Quantum Computers, have been discovered via QWs, e.g.
\cite{BooleanEvalQW,ConductivityQW}, or are better expressed using
QWs, e.g the Grover search. In these QW-based algorithms, the walker
usually explores a graph, which is encoding the instance of the problem.
No continuum limit is taken.\\
 The other reason is that a whole series of novel Quantum Simulation
schemes, for the near-future simulation devices, have been discovered
via QWs, and are better expressed as QWs \cite{Bialynicki-Birula,MeyerQLGI}.
Recall that quantum simulation is what motivated Feynman to introduce
the concept of Quantum Computing in the first place \cite{FeynmanQC}.
Whilst an universal Quantum Computer remains out-of-reach experimentally,
more special-purpose Quantum Simulation devices are seeing the light,
whose architecture in fact often resembles that of a QW \cite{WernerElectricQW,Sciarrino}.
In these QW-based schemes, the walker propagates on the square lattice,
and a continuum limit is taken to show that this converges towards
some well-known physics equation that one wishes to simulate. As an
added bonus, QW-based schemes provide: 1/ stable numerical schemes,
even for classical computers\textemdash thereby guaranteeing convergence
as soon as they are consistent \cite{ArrighiDirac}; 2/ simple discrete
toy models of the physical phenomena, that conserve most symmetries
(unitarity, homogeneity, causality, sometimes even Lorentz-covariance
\cite{arrighi2014discrete,DArianoLorentz}\textemdash thereby providing
playgrounds to discuss foundational questions in Physics \cite{LloydQG}.
It seems that QWs are unraveling as a new language to express quantum
physical phenomena.\\
 Whilst the present work is clearly within the latter trend, technically
it borrows from the former. Indeed, the QW-based schemes that we describe
depart from the square lattice, to go to the honeycomb and triangular
lattice\textemdash which opens the way for QW-based simulation schemes
on trivalent graphs.

{\em Motivations.} A motivation for this work is the possibility
to describe and implement the quantum simulation of certain physical
systems, without the need to rely on the square lattice architecture.
Rather, one would like to phrase a quantum simulation scheme in terms
of naturally occurring lattices in well-controlled substrates. Examples
of this class are the simulation of condensed matter systems modeled
by a tight-binding Hamiltonian, such as graphene \cite{neto2009electronic}
or the Kagome lattices \cite{Ye2018}\textemdash where the dynamics
of electrons can be effectively recast as a Dirac-like equation. In
fact the QW introduced in this paper may be useful as a simple point
of departure to predict electronic transport properties in the graphene
like-materials \cite{Bougroura2016} and exploring how varying their
geometry may influence the dispersion relations, and lead to topological
phases \cite{Kitagawa2010}, with interesting consequences on the
conducting properties.\\
 Another motivation for this work is to understand how fermions would
propagate if spacetime were a triangulated manifold, at the fundamental
level. Indeed, triangulated manifolds are being used to describe curved
spacetime since \cite{Regge1961}\textendash when Regge introduced
his simplicial, discrete formulation of General Relativity. This discrete
formulation then motivated a number of quantum gravity theories, such
as Loop Quantum Gravity \cite{Rovelli1998} and Causal Dynamical Triangulation
\cite{Ambjorn2006}\textemdash which seek to recover Regge calculus
in the classical limit. Most often quantum gravity research focuses
on the core issue of the quantum dynamics of discrete spacetime itself\textemdash overlooking
the question of how matter would propagate within the discrete spacetime
structure it prescribes. The present ideas may help address the question.

{\em Duality.} In a previous work, we showed how a QW can be defined
on the honeycomb and the triangular lattice \cite{PhysRevA.97.062111}
(see also \cite{Jay2018}), whose continuum limit is the Dirac equation
in $(2+1)$\textendash dimensional spacetime. Here, we extend these
definitions to allow for spacetime dependent local unitaries, and
introduce a dynamics that, in the continuum limit, corresponds to
the Dirac equation in a curved $(2+1)$\textendash dimensional spacetime.\\
 The construction, we feel, is interesting. Indeed, given a lattice
made of equilateral triangles, we begin by distorting the metric just
via a coordinate transformation, following the initial step of the
derivation of the Dirac equation in ordinary curved spacetime. But
then we realize that the coordinate transformation can be absorbed
by a suitable choice of the three gamma matrices that are associated
to the three directions provided by the triangles\textemdash a possibility
offered by the fact that these three directions are, of course, linearly-dependent
in the plane. Recall that the role of the gamma matrices is to prescribe
a basis of the spin, in which spin up goes one way, and spin down
goes the opposite way. In the QW, the local unitaries implement precisely
the corresponding changes of base. Thus, the gamma matrices determine
the local unitaries in the QW. This, therefore, unravels an equivalence,
in the continuum limit, between changing the actual geometry of the
lattice, or keeping it fixed but changing the local unitaries in a
suitable manner. The final step is to allow the local unitaries to
be spacetime dependent and take the continuum limit, thereby recovering
the Dirac equation in curved spacetime.\\
 Notice that having three directions in two-dimensional space, as
in the honeycomb or triangular lattices, is what provides that extra
degree of freedom allowing for the transfer of the geometric distortions
into the local unitaries\textemdash the square lattice is too rigid
in this respect.

{\em Related works.} It is already well known that QW can simulate
the Dirac equation \cite{BenziSucci,Bialynicki-Birula,MeyerQLGI,LapitskiDellarPalpacelliSucci,DAriano,ArrighiDirac,chandrashekar2013two},
the Klein-Gordon equation \cite{IndiansDirac,ArrighiKG,MolfettaDebbasch}
and the Schrödinger equation \cite{StrauchShrodinger,LoveBoghosian}
and that they are a minimal setting in which to simulate particles
in some inhomogeneous background field \cite{cedzich2013propagation,di2014quantum,marquez2017fermion,di2016quantum,arnault2016quantum},
with the difficult topic of interactions initiated in \cite{meyer1997quantum,ahlbrecht2012molecular}.
Eventually, the systematic study of the impact inhomogeneous local
unitaries also gave rise to QW models of particles propagating in
curved spacetime. This line of research was initiated by a QW simulations
of the curved Dirac equation in $(1+1)-$dimensions, for synchronous
coordinates \cite{di2013quantum,di2014quantum}, and later extended
by \cite{ArrighiGRDirac} to any spacetime metrics, and generalized
to further spatial and spin dimensions in \cite{DebbaschWaves,ArrighiGRDirac3D}.
All of these models were on the square lattice: to the best our knowledge
no one had modeled fermionic transport over non-square lattices. The
present paper shows that over the honeycomb and triangular lattices
the problem becomes considerably simpler, and the solution elegant.\\
 In a recent work \cite{stegmann2016current}, quantum transport over
curved spacetime has been compared to electronic transport in deformed
graphene, where a pseudo-magnetic field emulates an effective curvature
in the tight-binding Hamiltonian (see also \cite{kerner2012bending}).
Back to the quantum computing side, the Grover quantum search algorithm
has been expressed as a QW on the honeycomb lattice in \cite{Abal2010}
(and also in \cite{Foulger2015} with continuous time). Again for
quantum algorithmic purposes, \cite{Karafyllidis2015} studies the
possibility to use graphene nanoribbons to implement quantum gates.

{\em Plan.} The paper is organized as follows. In Sect. II we remind
the reader of the basic concepts and notations surrounding the Dirac
equation in a curved spacetime, in $(3+1)$ and $(2+1)$\textemdash dimensions.
In Sect, III we revisit our earlier Dirac QW on a honeycomb and on
a triangular lattice, and why it worked. In Sect. IV we show how a
simple, homogeneous coordinate transformation impacts the continuum
limit of the Dirac QW. Sect. V shows the duality, i.e. how the coordinate
transformation can be absorbed into a choice of local unitaries. Sect.
VI contains our main result: a QW that reproduces the Dirac equation
with curvature in the continuum limit, both for the honeycomb and
for the triangular lattices. We use $\hbar=c=1$ units.

\section{Dirac equation in curved spacetime: a recap}

\subsection{$(3+1)$\textemdash dimensions}

In this Section we recall the basic properties of the Dirac equation
in curved spacetime. We refer the reader to \cite{Lawrie,Koke2016,Yepez2011}
for a review. We start by describing the case of a $(3+1)$\textendash dimensional
spacetime with coordinates $x^{\mu}$, $\mu=0,\dots4$, where $x^{0}$
is the time coordinate, and metric tensor $g_{\mu\nu}(x)$ in these
coordinates. At each point $x$, it is possible to introduce a set
of four vectors $\{{e_{\mu}}^{a}(x)/a,\mu=0,\dots4\}$, referred to
as the tetrad or vierbein, that locally diagonalizes the metric tensor
i.e., 
\begin{equation}
g_{\mu\nu}(x)={e_{\mu}}^{a}(x){e_{\nu}}^{b}(x)\eta_{ab}.\label{eq:tetraddiag}
\end{equation}
(here and thereafter, summation over repeated indices is assumed),
where $\eta_{ab}=\textrm{Diag}(1,-1,-1,-1)$. Notice that, given a
vierbein, one can obtain a new one, which would also satisfy Eq. (\ref{eq:tetraddiag}),
by performing an arbitrary Lorentz transformation. The inverse of
the vierbein is denoted ${e^{\mu}}_{a}$ (interchanged indices), satisfying
\begin{equation}
{e^{\mu}}_{a}(x){e_{\nu}}^{a}(x)=\delta_{\nu}^{\mu},\quad{e_{\mu}}^{a}(x){e^{\mu}}_{b}(x)=\delta_{b}^{a}.\label{vierbein_orthonormality_conditions}
\end{equation}
Using \eqref{eq:tetraddiag} and \eqref{vierbein_orthonormality_conditions},
one has 
\begin{equation}
g_{\mu\nu}(x){e^{\mu}}_{a}(x)\,{e^{\nu}}_{b}(x)=\eta_{ab}.\label{eq:gwitheeqeta}
\end{equation}
Thus, tetrads can be understood as normalized tangent vectors that
relate the original coordinates to a local inertial frame. We use
the common convention that inertial coordinates are designated by
latin indices, and original coordinates by greek indices. Latin indices
are lowered and raised by $\eta_{ab}$, greek indices by $g_{\mu\nu}$.
In the local inertial frame, one is legitimated to use the Dirac $\gamma$\textendash matrices,
i.e. matrices satisfying the Clifford algebra $\{\gamma^{a},\gamma^{b}\}=\!2\eta^{ab}\boldsymbol{\mathbb{I}}$.
From these, one defines $\sigma^{ab}=\frac{i}{2}[\gamma^{a},\gamma^{b}]$.

Given a Dirac field $\psi(x)$, the action of a local Lorentz transformation
${\Lambda^{a}}_{b}(x)$ can be written as 
\begin{equation}
\psi\rightarrow U_{\Lambda}\psi,
\end{equation}
where 
\begin{equation}
U_{\Lambda}(x)=e^{-\frac{i}{4}\theta_{ab}(x)\sigma^{ab}},
\end{equation}

and $\theta_{ab}(x)$ are the parameters of the transformation, defined
by ${\Lambda^{a}}_{b}(x)=\delta_{b}^{a}+{\theta^{a}}_{b}(x)$. One
can prove that this operator acts on Dirac gamma matrices as follows:
\begin{equation}
U_{\Lambda}{}^{-1}\gamma^{a}U_{\Lambda}={\Lambda^{a}}_{b}\gamma^{b}.\label{eq:Lambda_D_similarity_transform}
\end{equation}
With the above notations, the Dirac equation in curved space 
\begin{equation}
i\gamma^{a}{e^{\mu}}_{a}(x){\cal D}_{\mu}\,\psi-m\,\psi=0,\label{eq:Dirac3plus1}
\end{equation}
where $m$ is the particle mass, is invariant under a local Lorentz
transformation provided the generalized derivative that we use is
\begin{equation}
{\cal D}_{\mu}=\partial_{\mu}+\Gamma_{\mu},
\end{equation}
where $\Gamma_{\mu}$ transforms according to 
\begin{equation}
\Gamma_{\nu}\longrightarrow\Gamma'_{\nu}=U_{\Lambda}\Gamma_{\nu}U_{\Lambda}{}^{-1}-\partial_{\nu}\left(U_{\Lambda}\right)U_{\Lambda}{}^{-1}.\label{Gamma_pseudo_gauge_transformation}
\end{equation}

The correction $\Gamma_{\mu}$ to the derivative can then be obtained
as \cite{Koke2016} 
\begin{align}
\Gamma_{\mu}(x)=-\frac{i}{4}\omega_{ab\mu}(x)\sigma^{ab},
\end{align}
where $\omega_{ab\mu}(x)$ is the so-called spin connection, and can
be expressed in terms of the tetrads and the affine connection as
\begin{equation}
\omega^{a}{}_{b\nu}={e_{\mu}}^{a}\partial_{\nu}{e^{\mu}}_{b}+{e_{\mu}}^{a}{e^{\sigma}}_{b}\Gamma_{\sigma\nu}^{\mu}.
\end{equation}
From Eq. (\ref{eq:Dirac3plus1}) one can define a four-vector current
\begin{equation}
j^{\mu}=\sqrt{g}{e^{\mu}}_{a}\bar{\psi}\gamma^{a}\psi,
\end{equation}
where $g$ is the (absolute value of) the determinant of the metric,
so that it is conserved: 
\begin{equation}
\partial_{\mu}j^{\mu}=0.
\end{equation}
This justifies the normalization condition 
\begin{equation}
\int j^{0}dv=\int\sqrt{g}{e^{0}}_{0}\psi^{\dagger}\psi dv=1,\label{eq:contnormalization}
\end{equation}
with $dv$ the volume element in space.

\subsection{$(2+1)$\textemdash dimensions}

When the space dimension is lower than $3$, the $\gamma$\textendash matrices
become $2\times2$. Then, the Dirac Eq. (\ref{eq:Dirac3plus1}) can
be simplified to give 
\begin{equation}
i\boldsymbol{\gamma}^{a}\left[{e^{\mu}}_{a}\partial_{\mu}\psi\!+\!\frac{1}{2\sqrt{g}}\partial_{\mu}\left({e^{\mu}}_{a}\sqrt{g}\right)\psi\right]-m\psi=0.\label{eq:CDE}
\end{equation}
We will now express this equation in Hamiltonian form. We name the
greek indices $\mu=t,x,y$, and the latin indices $a=0,1,2$. By performing
a local Lorentz transformation, it is possible to arrive to a form
of the tetrad such that ${e^{t}}_{a}=0$ for $a=1,2$. Then, by introducing
the change of wavefunction given by \cite{de1962representations}:
\begin{equation}
\chi=g^{1/4}({e^{t}}_{0})^{1/2}\psi\label{eq:psitochi}
\end{equation}
and multiplying Eq. (\ref{eq:CDE}) by $\beta\equiv\gamma^{0}$, one
gets 
\begin{equation}
i\partial_{t}\chi+\frac{i}{2}\{B^{s},\partial_{s}\}\chi-\frac{m}{{e^{t}}_{0}}\beta\chi=0,\label{eq:Oliveira}
\end{equation}
where $s=1,2$, and we have introduced the notation \textbf{$B^{s}=\alpha^{a}\frac{{e^{s}}_{a}}{{e^{t}}_{0}}$},
with the usual Dirac $\alpha$\textendash matrices $\alpha^{a}\equiv\beta\gamma^{a}$.
In particular, one can make the choice $\gamma^{0}=\sigma^{z}$,$\gamma^{1}=i\sigma^{y}$
and $\gamma^{2}=-i\sigma^{x}$. Then $\alpha^{0}$ becomes the identity
matrix, $\alpha^{1}=\sigma^{x}$ and $\alpha^{2}=\sigma^{y}$, with
$\sigma^{i}$ ($i=1,2,3$) the Pauli matrices.

According to Eqs. (\ref{eq:contnormalization}) and (\ref{eq:psitochi}),
the normalization condition becomes simply 
\begin{equation}
\int\chi^{\dagger}\chi dv=1.
\end{equation}

\section{Dirac QW}

\noindent A possible representation of the Dirac equation in flat
spacetime is obtained from Eq. (\ref{eq:Oliveira}) by using the canonical
tetrads ${e^{\mu}}_{a}=\delta_{a}^{\mu}$ and the choice of Dirac
$\alpha$\textendash matrices made at the end of Sect. II: 
\begin{align}
i\partial_{t}\psi & =H_{D}\psi\quad\textrm{with}\quad H_{D}=p_{x}\sigma^{x}+p_{y}\sigma^{y}+m\sigma^{z}.\label{eq:Dirac2D}
\end{align}
where $p_{i}$is the $i^{\textrm{th}}$ component of the momentum
operator.

It is now very well-known that one can define a QW on the lattice
that converges, in the limit of both the lattice spacing and the time
step going to zero, towards the solutions of \eqref{eq:Dirac2D}.
This is done by defining a Hilbert space $\mathcal{H}=\mathcal{H}_{x}\otimes\mathcal{H}_{y}\otimes\mathcal{H}_{c}$,
where $\mathcal{H}_{x}\otimes\mathcal{H}_{y}$ represents the space
degrees of freedom and is spanned by the basis states $\ket{x=\varepsilon j,y=\varepsilon k}$
with $j,k\in\mathbb{Z}$, whereas $\mathcal{H}_{c}=\textrm{Span}\{|c\rangle/c\in\{-1,1\}\}$
describes the internal `coin' (spin) degree of freedom. Over $\mathcal{H}_{x}\otimes\mathcal{H}_{y}$,
the $p_{i}$ will now denote the quasimomentum operators defined by
\begin{eqnarray}
\exp(-i\varepsilon p_{x})\ket{x,y} & = & \ket{x+\varepsilon,y}\nonumber \\
\exp(-i\varepsilon p_{y})\ket{x,y} & = & \ket{x,y+\varepsilon}.
\end{eqnarray}
The Dirac QW will evolve a state $\psi(t)$ into 
\begin{align}
\psi(t+\varepsilon) & =\exp(-im\varepsilon\sigma^{z})\exp(-i\varepsilon p_{x}\sigma^{x})\exp(-i\varepsilon p_{y}\sigma^{y})\nonumber \\
 & \approx\exp(-i\varepsilon H_{D})\psi(t)\label{eq:evolint}
\end{align}
using the Lie-Trotter formula. It follows that one recovers the Dirac
equation (\ref{eq:Dirac2D}) in the continuum limit when $\varepsilon$
goes to zero, where the $p_{i}$ become the true momentum operators
$p_{i}=-i\partial_{i}$.

Recently \cite{PhysRevA.97.062111} we showed that Dirac dynamics
can be implemented by a QW, not only over square lattices, but also
over the honeycomb and triangular lattices (see also \cite{Jay2018}).
The honeycomb lattice QW is easier to introduce. It defines three
directions $u_{i}$, $i=0,1,2$ having relative angles of $120^{\circ}$,
let $u_{i}^{j}$ denote their coordinates. The idea is to introduce
three unitary $2\times2$\textendash matrices $\tau^{i}$ with eigenvalues
$\pm1$ such that $H_{D}$ can be written as 
\begin{equation}
H_{D}=\pi_{i}\tau^{i}+m\sigma^{z},\label{eq:HDpitaus}
\end{equation}
where $\pi_{i}\equiv u_{i}^{j}p_{j}$ represents the quasimomentum
operator along the $u_{i}$ direction. Then, the corresponding QW
can again be defined by a Lie-Trotter expansion of Eq. (\ref{eq:evolint}),
with $H_{D}$ defined in (\ref{eq:HDpitaus}). The triangular lattice
QW makes use of a similar setup, although the translations are generated
by rotations of the triangles themselves, bringing apart the internal
components of the field $\psi$, which is assumed to `live' in the
edges of the triangles, one component ( $\psi^{\uparrow}$or $\psi^{\downarrow}$)
on each side.

\section{Coordinate transformation on the Dirac equation }

The construction of the Dirac equation in curved spacetime relies
on the equivalence principle, which means that one can introduce a
local transformation of coordinates at a given point, so that one
recovers the flat equation in the neighborhood of that point. The
curved Dirac equation is then that which stems from applying the reverse
the local tranformation, upon the flat Dirac equation. Our line of
thought follows that step, i.e., starting from the flat case Dirac
QW, perform an arbitrary change of coordinates so as to obtain the
curved Dirac QW. Let us begin with just an homogeneous change of coordinates
on the Dirac equation.

First notice that Eq. (\ref{eq:gwitheeqeta}) can be writen as $e^{T}ge=\eta$,
where $e$ is just the representation of the tetrads in matricial
form, and $^{T}$ denotes the matrix transpose. Now, under a global
change of coordinates $\Gamma$ such that $x'=\Gamma x$, the metric
$g$ and the vierbein transform as 
\begin{align}
g & \ \mapsto g'=({\Gamma^{T}})^{-1}g\Gamma^{-1}\nonumber \\
e & \ \mapsto e'=\Gamma e\label{eq:transfeg}
\end{align}
This transformation fulfills the tetrads-metric relation, 
\begin{equation}
{e'}^{T}g'{e}'=e^{T}\Gamma^{T}({\Gamma^{T}})^{-1}g\Gamma^{-1}{\Gamma}e=e^{T}ge=\eta.
\end{equation}
Next we start from a QW that reproduces the flat equation, and introduce
a deformation (described by the transformation $\Gamma$) that will
end up with a more generic metric $g'$. We can make a simple choice,
given by the canonical tetrads ${e^{\mu}}_{a}=\delta_{a}^{\mu}$ for
the initial coordinates, and then transform them according to Eq.
(\ref{eq:transfeg}). Since we are considering a deformation of the
spatial sites of the lattice, the time components will be left unchanged,
and the matrix $\Gamma$ will take the form 
\begin{equation}
\Gamma=\left(\begin{array}{ccc}
1 & 0 & 0\\
0 & \lambda_{11} & \lambda_{12}\\
0 & \lambda_{21} & \lambda_{22}
\end{array}\right).
\end{equation}
where each $\lambda_{ij}$ are position independent, although they
are allowed to depend on time.

Under this restriction, we can reduce the problem to a transformation
on a bidimensional space, where ${e^{t}}_{0}=1$, which implies that
Eq. (\ref{eq:Oliveira}) adopts the simpler form 
\begin{equation}
i\partial_{t}\chi+\frac{i}{2}\{B^{s},\partial_{s}\}\chi-m\beta\chi=0.\label{eq:Oliveirae001}
\end{equation}
Let us consider how this transformation will affect the QW defined
on a triangular lattice, as introduced in Sect. III (see \cite{PhysRevA.97.062111}).
Such transformation will imply modifying the vectors $u_{i}$, yielding
the new vectors 
\begin{equation}
u'_{i}=\left(\begin{array}{cc}
\lambda_{11} & \lambda_{12}\\
\lambda_{21} & \lambda_{22}
\end{array}\right)u_{i}\equiv\Lambda u_{i}.\label{eq:transu}
\end{equation}

Introducing these vectors in our algorithms and calculating the continuum
limit, we arrive to the following equation \begin{small} 
\begin{equation}
i\partial_{t}\psi=\left[\left(\lambda_{11}\sigma^{x}+\lambda_{12}\sigma^{y}\right)p_{x}+\left(\lambda_{21}\sigma^{x}+\lambda_{22}\sigma^{y}\right)p_{y}\right]\psi+m\sigma^{z}\psi,\label{eq:diracmod}
\end{equation}
\end{small} which describes the Dirac equation on a flat geometry.
A comparison with Eq. (\ref{eq:Oliveira}) gives 
\begin{align}
B^{x} & =\lambda_{11}\sigma^{x}+\lambda_{12}\sigma^{y}\label{eq:Bsfromsigmasflat1}\\
B^{y} & =\lambda_{21}\sigma^{x}+\lambda_{22}\sigma^{y}.\label{eq:Bsfromsigmasflat2}
\end{align}
This procedure can be used for an homogeneous transformation, such
as the one defined above. In the next section, we introduce an alternative,
which consists in redefining the $\tau^{i}$ matrices. As we shall
see, this redefinition also allows for an inhomogeneous (i.e., space-time
dependent) $\Lambda(t,x,y)$ transformation, thereby resulting in
a Dirac equation in curved space.

\section{Curved Dirac equation from a non-homogeneous QW}

We now generalize the ideas developed in the previous Sect. with the
purpose to obtain, in the continuum limit, the Dirac equation on a
curved spacetime, for a given metrics with a triangular tetrad, as
discussed in Sect. II. We start by looking at the set of matrices
\textbf{$B^{s}=\alpha^{a}\frac{{e^{s}}_{a}}{{e^{t}}_{0}}$}, as a
linear transformation over the set of usual Pauli matrices, in the
same spirit as Eqs. (\ref{eq:Bsfromsigmasflat1}) and (\ref{eq:Bsfromsigmasflat2}).
This leads us to define the transformation $\Lambda(t,x,y)$, with
matrix elements 
\begin{equation}
\Lambda_{a}^{s}\equiv\frac{{e^{s}}_{a}}{{e^{t}}_{0}}
\end{equation}
(we have omitted the time and space dependence for convenience). Then,
the above mentioned transformation reads 
\begin{equation}
B^{s}=\Lambda_{a}^{s}\alpha^{a}.
\end{equation}

We now make use of the property that relates the $\tau^{i}$ matrices,
defined in Eq. (\ref{eq:HDpitaus}), with the Pauli matrices: $u_{i}^{k}\tau^{i}=\sigma^{k}$
(see \cite{PhysRevA.97.062111}). In this way, we arrive to 
\begin{equation}
B^{s}=\Lambda_{k}^{s}u_{i}^{k}\tau^{i}.
\end{equation}

The above equation can be understood as a transformation performed
on the $u_{i}$ vectors, c.f. Eq. (\ref{eq:transu}), as the origin
of the curved spacetime equation.

Instead of introducing a distortion $\Lambda(t,x,y)$ on the lattice
via the modification of the $u_{i}$ vectors, the unitary matrices
$\tau^{i}$ can be transformed to produce the same effect. In other
words, we seek for a set of matrices $\beta^{i}(t,x,y)$ that fulfill
the following conditions: 
\begin{itemize}
\item (C1) We impose that 
\begin{equation}
\Lambda_{k}^{j}(t,x,y)u_{i}^{k}\tau^{i}=u_{i}^{j}\beta^{i}(t,x,y).\label{eq:sumpi}
\end{equation}
\item (C2) Each of them has $\{-1,1\}$ as eigenvalues, i.e. at any time
step and at any point $(x,y)$ of the lattice there exist three unitaries
$U_{i}(t,x,y)$ such that 
\begin{equation}
\beta^{i}(t,x,y)=U_{i}^{\dagger}(t,x,y)\sigma^{z}U_{i}(t,x,y).\label{ref:beta}
\end{equation}
\end{itemize}
Notice that condition (C1) implies that the coordinate transformation
dictated by $\Lambda_{k}^{j}(t,x,y)$ is transferred to the unitary
operations, which become new spacetime dependent $\beta^{i}(t,x,y)$,
instead of the original $\tau^{i}$. Additionally, condition (C2)
will allow us to rewrite the QW evolution in terms of the usual state-dependent
translation operators. Let us apply these ideas to the honeycomb and
the triangular lattice.

To alleviate the notations, in what follows we will omit the spacetime
dependence both in these matrices and in the $U_{i}(t,x,y)$, and
write simply $\beta^{i}$ and $U_{i}$. The above conditions allow
to calculate the $\beta^{i}$ matrices, which can be written as a
combination of Pauli matrices, i.e. $\beta^{i}=\vec{n}^{i}\cdot\vec{\sigma}$,
where each $\vec{n}^{i}$ must be a real, unit vector $\vec{n}^{i}=(\sin{\theta_{i}}\cos{\phi_{i}},\sin\theta_{i}\sin\phi_{i},\cos\theta_{i})$
for some angles $\theta_{i}$ and $\phi_{i}$ (that are time and position
dependent).

In this way 
\begin{equation}
\beta_{i}=U_{i}^{\dagger}\sigma_{z}U_{i}=\left(\begin{matrix}\cos{\theta_{i}} &  & e^{-i\phi_{i}}\sin{\theta_{i}}\\
e^{i\phi_{i}}\sin{\theta_{i}} &  & -\cos{\theta_{i}}
\end{matrix}\right),
\end{equation}
and each $U_{i}$ can be obtained by diagonalization of the corresponding
$\beta^{i}$. With an appropriate choice of phases, we finally write
them as 
\begin{equation}
U_{i}=\left(\begin{matrix}e^{\frac{i\phi_{i}}{2}}\cos{\frac{\theta_{i}}{2}} &  & e^{-\frac{i\phi_{i}}{2}}\sin{\frac{\theta_{i}}{2}}\\
-e^{\frac{i\phi_{i}}{2}}\sin{\frac{\theta_{i}}{2}} &  & e^{-\frac{i\phi_{i}}{2}}\cos{\frac{\theta_{i}}{2}},
\end{matrix}\right).\label{eq:choiceUis}
\end{equation}

Before we proceed to examine the induced QW on the honeycomb and triangular
lattices together with their limits, let us discuss what the situation
would have been in the square lattice, had we implement the above
procedure. In this case, the original Dirac matrices can be chosen
to be the Pauli matrices, and the two unit vectors $u_{i}$ can be
taken to be the canonical ones, so that the requirement of Eq. (\ref{eq:sumpi})
simply becomes 
\begin{equation}
\Lambda_{k}^{j}\sigma^{k}=\beta^{j}.\label{eq:dualitysquare}
\end{equation}
But then, since condition (C2) implies that $\det(\beta^{j})=-1$
for each $j$, we need that 
\begin{equation}
\sum_{k}(\Lambda_{k}^{j})^{2}=1.\label{eq:conditionlambdassquare}
\end{equation}
Thus the square lattice only allows for a limited form of ``duality'',
i.e. only those transformations satisfying condition (\ref{eq:conditionlambdassquare})
can be absorbed into the unitaries, whereas the honeycomb and triangular
lattices allow for arbitrary transformations.

\section{Curved Dirac QW}

\subsection{Honeycomb QW}

In this section we define the QW over the honeycomb, following a similar
procedure as in \cite{PhysRevA.97.062111}. After the ideas developed
in Sect. V, we define the following Hamiltonian to be used in the
QW: 
\begin{equation}
\mathcal{H}=\frac{1}{2}u_{i}^{j}\left(\beta^{i}p_{j}+p_{j}\beta^{i}\right)+\widetilde{m}\sigma^{z}\label{eq:Hcurved}
\end{equation}
with $\widetilde{m}=m/{e^{t}}_{0}$. Expanding the Hamiltonian, we
arrive to: 
\begin{equation}
\mathcal{H}=-iu_{i}^{j}U_{i}^{\dagger}\sigma_{z}\partial_{j}U_{i}-\frac{i}{2}u_{i}^{j}\left[(\partial_{j}U_{i}^{\dagger})\sigma_{z}U_{i}-U_{i}^{\dagger}\sigma_{z}(\partial_{j}U_{i})\right]+\widetilde{m}\sigma^{z}
\end{equation}
After substitution of Eq. (\ref{eq:choiceUis}), one obtains 
\begin{equation}
(\partial_{j}U_{i}^{\dagger})\sigma_{z}U_{i}-U_{i}^{\dagger}\sigma_{z}(\partial_{j}U_{i})=-i\cos\theta_{i}\partial_{j}\phi_{j}\mathbb{I},\label{eq:diffderUis}
\end{equation}
with $\mathbb{I}$ the identity matrix. Notice that, unlike in the
flat space situation, there is no possible choice of the phases in
the $U_{i}$s that makes Eq. (\ref{eq:diffderUis}) vanish for all
values of $i$. One may wonder whether there is a reason behind this,
for example the existence of some topological or gauge invariant that
forbids all these quantities to be simultaneously zero. This issue
might deserve further investigation in the future. In any case, the
additional term in Eq. (\ref{eq:diffderUis}) that arises from the
choice given by Eq. (\ref{eq:choiceUis}) contributes only as a space-time
dependent phase, which is easy to handle both from the theoretical
and from the experimental point of view. We finally arrive to: 
\begin{equation}
\mathcal{H}=\sum_{i}\left(U_{i}^{\dagger}\sigma_{z}\pi_{i}U_{i}+\gamma_{i}\mathbb{I}\right)+\widetilde{m}\sigma^{z}
\end{equation}
where $\gamma_{i}=-\frac{i}{2}\cos\theta_{i}\pi_{i}\phi_{i}$. In
order to define the QW, we make use of the Lie-Trotter product formula
to decompose the evolution of the wavefunction $\psi(t+\epsilon)=e^{-i\epsilon\mathcal{H}}\psi(t)$
as a product of unitary matrices 
\begin{align}
e^{-i\epsilon\left[\sum_{i}\left(U_{i}^{\dagger}\sigma_{z}\pi_{i}U_{i}+\gamma_{i}\right)+\widetilde{m}\sigma^{z}\right]} & \approx\nonumber \\
e^{-i\widetilde{m}\varepsilon\sigma^{z}}\prod_{i}e^{-i\epsilon U_{i}^{\dagger}\sigma_{z}\pi_{i}U_{i}}e^{-i\epsilon\gamma_{i}}.
\end{align}
Applying condition (C1), and introducing the translation operators
along the $u_{i}$ direction as $T_{i}=e^{-i\epsilon\sigma^{z}\pi_{i}}$,
the QW on a honeycomb can be defined as: 
\begin{equation}
\psi(t+\epsilon)=e^{-i\widetilde{m}\varepsilon\sigma^{z}}\prod_{i}U{^{\dagger}}_{i}T_{i}U_{i}e^{-i\epsilon\gamma_{i}}
\end{equation}

By construction, in the continuous limit, we arrive to the Dirac equation
in 2+1 curved space-time, under the form 
\begin{equation}
i\partial_{t}\psi=\frac{1}{2}\left[u_{i}^{j}\beta^{i}(t,x,y)p_{j}+u_{i}^{j}p_{j}\beta^{i}(t,x,y)\right]\psi+\widetilde{m}\sigma^{z}\psi.\label{eq:betaeq}
\end{equation}
As expected, this equation can be nicely rewritten under the form
Eq. (\ref{eq:Oliveira}), if we define $B^{j}(t,x,y)\equiv u_{i}^{j}\beta^{i}(t,x,y)$.

\subsection{Triangular QW}

Let us describe first the dynamics corresponding to the massless case.
Again, we follow the same procedure as in \cite{PhysRevA.97.062111}.
The triangles are equilateral, with sides labeled by $k=0,1,2$. The
two-dimensional spinors are assumed to lie on the edges shared by
neighboring triangles. We denote them by $\psi(t,v,k)=\left(\begin{array}{c}
\psi^{\uparrow}(t,v,k)\\
\psi^{\downarrow}(t,v,k)
\end{array}\right)$, with $v$ a triangle and $k$ a side. Therefore, the position at
the lattice will be labeled by $(v,k)$. The dynamics of the Triangular
QW is defined as the composition of three operators. The first operator
consists on the application of the $2\times2$ unitary matrix $U_{i}(t,v,k)$,
defined in the last section, to each two-dimensional spinor on every
edge shared by two neighboring triangles. The second operator, $R$,
simply rotates every triangle anti-clockwise. The third operator is
just the application of the unitary matrix $U_{i}^{\dagger}(t,v,k+1)$
again at each edge shared by two neighboring triangles, where the
addition $k+1$ is understood modulo $2$. Altogether, the Triangular
QW dynamics is given by: 
\begin{widetext}
\begin{eqnarray}
\psi(t+\varepsilon/3,v,k)=U_{i}^{\dagger}(t,v,k)\left[P^{\uparrow}U_{i}(t,v,k-1)e^{-i\epsilon\gamma_{i}}\right.\psi(t,v,k-1)\nonumber \\
\oplus\left.P^{\downarrow}U_{i}(t,e(v,k),k-1)e^{-i\epsilon\gamma_{i}}\psi(t,e(v,k),k-1)\right]\equiv W_{i}(t)\psi(t)
\end{eqnarray}
where $P^{\uparrow}$ and $P^{\downarrow}$ are the projectors over
the upper and lower component of the spinor, respectively, and $e(t,v,k)$
is the neighbor of triangle $v$ alongside $k$ at fixed time t. We
define one timestep of the evolution by the composition of the three
operators $W_{i}$, and include the mass term, as follows 
\begin{eqnarray}
\psi(t+\varepsilon)=e^{-i\widetilde{m}\varepsilon\sigma^{z}}(W_{2}W_{1}W_{0})\psi(t)
\end{eqnarray}
\end{widetext}

By expanding this equation up to first order in $\varepsilon$, after
a tedious but straightforward computation, one arrives to the following
equation in the continuum limit: 
\begin{widetext}
\begin{eqnarray}
\partial_{t}\psi=(U_{0}^{\dagger}\sigma^{z}U_{0}-\frac{1}{2}U_{1}^{\dagger}\sigma^{z}U_{1}-\frac{1}{2}U_{2}^{\dagger}\sigma^{z}U_{2})\partial_{x}\psi+\frac{\sqrt{3}}{2}(U_{1}^{\dagger}\sigma^{z}U_{1}-U_{2}^{\dagger}\sigma^{z}U_{2})\partial_{y}\psi+\nonumber \\
\partial_{x}(U_{0}^{\dagger}\sigma^{z}U_{0}-\frac{1}{2}U_{1}^{\dagger}\sigma^{z}U_{1}-\frac{1}{2}U_{2}^{\dagger}\sigma^{z}U_{2})\psi+\frac{\sqrt{3}}{2}\partial_{y}(U_{1}^{\dagger}\sigma^{z}U_{1}-U_{2}^{\dagger}\sigma^{z}U_{2})\psi-i\widetilde{m}\sigma^{z}\psi\label{eq:continuumtriang}
\end{eqnarray}
\end{widetext}

where the above terms appear from an expansion at order $O(\varepsilon)$.

Notice that, if we define $B^{x}\equiv(\beta^{0}-\frac{1}{2}\beta^{1}-\frac{1}{2}\beta^{2})$,
and $B^{y}\equiv\frac{\sqrt{3}}{2}(\beta\text{\textonesuperior}-\beta\text{\texttwosuperior})$,
Eq. (\ref{eq:continuumtriang}) adopts the desired form of (\ref{eq:Oliveira}).

\section{Discussion}

We introduced a Quantum Walk (QW) over the honeycomb and the triangular
lattice. In both cases, our starting point was the possibility to
rewrite the targeted Hamiltonian as a sum of momentum operators along
the three relevant directions of the lattice, each weighted by a suitably
chosen gamma matrix. This procedure has been introduced in \cite{PhysRevA.97.062111}\textemdash our
targeted Hamiltonian was then that of the Dirac equation, which we
recovered in the continuum limit. In the present work, we realized
that due to the linear dependence of the three preferred directions
of the honeycomb and the triangular lattices, one could also obtain
the Hamiltonian of the Dirac equation under an arbitrary change of
coordinates. We emphasized that applying the same procedure, but for
the square lattice, only allows for a very limited set of changes
of coordinates.\\
 Then, by making the gamma matrices to be spacetime dependent, we
obtained the Curved Dirac equation in an arbitrary background metric.
Overall, the QW hereby constructed over the honeycomb and the triangular
lattices thus recovers, in the continuum limit, the Dirac equation
in curved $(2+1)$\textendash dimensional spacetime. We believe that
the duality between changes of metric, and changes of gamma matrices
weighting non linearly-independent momentum operators, is profound
and may lead to further developments. 
\begin{acknowledgments}
\noindent We acknowledge the very enlightening discussion on general
covariance with Luca Fabbri. This work has been funded by the INFINITI
and the CNRS PEPs Spain-France PIC2017FR6, the STICAmSud project 16STIC05
FoQCoSS and the Spanish Ministerio de Economía, Industria y Competitividad
, MINECO-FEDER project FPA2017-84543-P, SEV-2014-0398 and Generalitat
Valenciana grant GVPROMETEOII2014-087. 
\end{acknowledgments}

 \bibliographystyle{apsrev4-1}

\begin{thebibliography}{49}%
\makeatletter
\providecommand \@ifxundefined [1]{%
 \@ifx{#1\undefined}
}%
\providecommand \@ifnum [1]{%
 \ifnum #1\expandafter \@firstoftwo
 \else \expandafter \@secondoftwo
 \fi
}%
\providecommand \@ifx [1]{%
 \ifx #1\expandafter \@firstoftwo
 \else \expandafter \@secondoftwo
 \fi
}%
\providecommand \natexlab [1]{#1}%
\providecommand \enquote  [1]{``#1''}%
\providecommand \bibnamefont  [1]{#1}%
\providecommand \bibfnamefont [1]{#1}%
\providecommand \citenamefont [1]{#1}%
\providecommand \href@noop [0]{\@secondoftwo}%
\providecommand \href [0]{\begingroup \@sanitize@url \@href}%
\providecommand \@href[1]{\@@startlink{#1}\@@href}%
\providecommand \@@href[1]{\endgroup#1\@@endlink}%
\providecommand \@sanitize@url [0]{\catcode `\\12\catcode `\$12\catcode
  `\&12\catcode `\#12\catcode `\^12\catcode `\_12\catcode `\%12\relax}%
\providecommand \@@startlink[1]{}%
\providecommand \@@endlink[0]{}%
\providecommand \url  [0]{\begingroup\@sanitize@url \@url }%
\providecommand \@url [1]{\endgroup\@href {#1}{\urlprefix }}%
\providecommand \urlprefix  [0]{URL }%
\providecommand \Eprint [0]{\href }%
\providecommand \doibase [0]{http://dx.doi.org/}%
\providecommand \selectlanguage [0]{\@gobble}%
\providecommand \bibinfo  [0]{\@secondoftwo}%
\providecommand \bibfield  [0]{\@secondoftwo}%
\providecommand \translation [1]{[#1]}%
\providecommand \BibitemOpen [0]{}%
\providecommand \bibitemStop [0]{}%
\providecommand \bibitemNoStop [0]{.\EOS\space}%
\providecommand \EOS [0]{\spacefactor3000\relax}%
\providecommand \BibitemShut  [1]{\csname bibitem#1\endcsname}%
\let\auto@bib@innerbib\@empty
\bibitem [{\citenamefont {Arrighi}\ \emph {et~al.}(2018)\citenamefont
  {Arrighi}, \citenamefont {Di~Molfetta}, \citenamefont
  {M\'arquez-Mart\'{\i}n},\ and\ \citenamefont {P\'erez}}]{PhysRevA.97.062111}%
  \BibitemOpen
  \bibfield  {author} {\bibinfo {author} {\bibfnamefont {P.}~\bibnamefont
  {Arrighi}}, \bibinfo {author} {\bibfnamefont {G.}~\bibnamefont
  {Di~Molfetta}}, \bibinfo {author} {\bibfnamefont {I.}~\bibnamefont
  {M\'arquez-Mart\'{\i}n}}, \ and\ \bibinfo {author} {\bibfnamefont
  {A.}~\bibnamefont {P\'erez}},\ }\href {\doibase 10.1103/PhysRevA.97.062111}
  {\bibfield  {journal} {\bibinfo  {journal} {Phys. Rev. A}\ }\textbf {\bibinfo
  {volume} {97}},\ \bibinfo {pages} {062111} (\bibinfo {year}
  {2018})}\BibitemShut {NoStop}%
\bibitem [{\citenamefont {Ambainis}\ \emph {et~al.}(2010)\citenamefont
  {Ambainis}, \citenamefont {Childs}, \citenamefont {Reichardt}, \citenamefont
  {{\v{S}}palek},\ and\ \citenamefont {Zhang}}]{BooleanEvalQW}%
  \BibitemOpen
  \bibfield  {author} {\bibinfo {author} {\bibfnamefont {A.}~\bibnamefont
  {Ambainis}}, \bibinfo {author} {\bibfnamefont {A.~M.}\ \bibnamefont
  {Childs}}, \bibinfo {author} {\bibfnamefont {B.~W.}\ \bibnamefont
  {Reichardt}}, \bibinfo {author} {\bibfnamefont {R.}~\bibnamefont
  {{\v{S}}palek}}, \ and\ \bibinfo {author} {\bibfnamefont {S.}~\bibnamefont
  {Zhang}},\ }\href@noop {} {\bibfield  {journal} {\bibinfo  {journal} {SIAM
  Journal on Computing}\ }\textbf {\bibinfo {volume} {39}},\ \bibinfo {pages}
  {2513} (\bibinfo {year} {2010})}\BibitemShut {NoStop}%
\bibitem [{\citenamefont {Wang}(2017)}]{ConductivityQW}%
  \BibitemOpen
  \bibfield  {author} {\bibinfo {author} {\bibfnamefont {G.}~\bibnamefont
  {Wang}},\ }\href {http://dl.acm.org/citation.cfm?id=3179568.3179573}
  {\bibfield  {journal} {\bibinfo  {journal} {Quantum Info. Comput.}\ }\textbf
  {\bibinfo {volume} {17}},\ \bibinfo {pages} {987} (\bibinfo {year}
  {2017})}\BibitemShut {NoStop}%
\bibitem [{\citenamefont {Bialynicki-Birula}(1994)}]{Bialynicki-Birula}%
  \BibitemOpen
  \bibfield  {author} {\bibinfo {author} {\bibfnamefont {I.}~\bibnamefont
  {Bialynicki-Birula}},\ }\href@noop {} {\bibfield  {journal} {\bibinfo
  {journal} {Phys. Rev. D.}\ }\textbf {\bibinfo {volume} {49}},\ \bibinfo
  {pages} {6920} (\bibinfo {year} {1994})}\BibitemShut {NoStop}%
\bibitem [{\citenamefont {Meyer}(1996)}]{MeyerQLGI}%
  \BibitemOpen
  \bibfield  {author} {\bibinfo {author} {\bibfnamefont {D.~A.}\ \bibnamefont
  {Meyer}},\ }\href@noop {} {\bibfield  {journal} {\bibinfo  {journal} {J.
  Stat. Phys}\ }\textbf {\bibinfo {volume} {85}},\ \bibinfo {pages} {551}
  (\bibinfo {year} {1996})}\BibitemShut {NoStop}%
\bibitem [{\citenamefont {Feynman}(1982)}]{FeynmanQC}%
  \BibitemOpen
  \bibfield  {author} {\bibinfo {author} {\bibfnamefont {R.~P.}\ \bibnamefont
  {Feynman}},\ }\href@noop {} {\bibfield  {journal} {\bibinfo  {journal}
  {International Journal of Theoretical Physics}\ }\textbf {\bibinfo {volume}
  {21}},\ \bibinfo {pages} {467} (\bibinfo {year} {1982})}\BibitemShut
  {NoStop}%
\bibitem [{\citenamefont {Genske}\ \emph {et~al.}(2013)\citenamefont {Genske},
  \citenamefont {Alt}, \citenamefont {Steffen}, \citenamefont {Werner},
  \citenamefont {Werner}, \citenamefont {Meschede},\ and\ \citenamefont
  {Alberti}}]{WernerElectricQW}%
  \BibitemOpen
  \bibfield  {author} {\bibinfo {author} {\bibfnamefont {M.}~\bibnamefont
  {Genske}}, \bibinfo {author} {\bibfnamefont {W.}~\bibnamefont {Alt}},
  \bibinfo {author} {\bibfnamefont {A.}~\bibnamefont {Steffen}}, \bibinfo
  {author} {\bibfnamefont {A.~H.}\ \bibnamefont {Werner}}, \bibinfo {author}
  {\bibfnamefont {R.~F.}\ \bibnamefont {Werner}}, \bibinfo {author}
  {\bibfnamefont {D.}~\bibnamefont {Meschede}}, \ and\ \bibinfo {author}
  {\bibfnamefont {A.}~\bibnamefont {Alberti}},\ }\href@noop {} {\bibfield
  {journal} {\bibinfo  {journal} {Physical review letters}\ }\textbf {\bibinfo
  {volume} {110}},\ \bibinfo {pages} {190601} (\bibinfo {year}
  {2013})}\BibitemShut {NoStop}%
\bibitem [{\citenamefont {Sansoni}\ \emph {et~al.}(2012)\citenamefont
  {Sansoni}, \citenamefont {Sciarrino}, \citenamefont {Vallone}, \citenamefont
  {Mataloni}, \citenamefont {Crespi}, \citenamefont {Ramponi},\ and\
  \citenamefont {Osellame}}]{Sciarrino}%
  \BibitemOpen
  \bibfield  {author} {\bibinfo {author} {\bibfnamefont {L.}~\bibnamefont
  {Sansoni}}, \bibinfo {author} {\bibfnamefont {F.}~\bibnamefont {Sciarrino}},
  \bibinfo {author} {\bibfnamefont {G.}~\bibnamefont {Vallone}}, \bibinfo
  {author} {\bibfnamefont {P.}~\bibnamefont {Mataloni}}, \bibinfo {author}
  {\bibfnamefont {A.}~\bibnamefont {Crespi}}, \bibinfo {author} {\bibfnamefont
  {R.}~\bibnamefont {Ramponi}}, \ and\ \bibinfo {author} {\bibfnamefont
  {R.}~\bibnamefont {Osellame}},\ }\href {\doibase
  10.1103/PhysRevLett.108.010502} {\bibfield  {journal} {\bibinfo  {journal}
  {Phys. Rev. Lett.}\ }\textbf {\bibinfo {volume} {108}},\ \bibinfo {pages}
  {010502} (\bibinfo {year} {2012})}\BibitemShut {NoStop}%
\bibitem [{\citenamefont {Arrighi}\ \emph {et~al.}(2013)\citenamefont
  {Arrighi}, \citenamefont {Forets},\ and\ \citenamefont
  {Nesme}}]{ArrighiDirac}%
  \BibitemOpen
  \bibfield  {author} {\bibinfo {author} {\bibfnamefont {P.}~\bibnamefont
  {Arrighi}}, \bibinfo {author} {\bibfnamefont {M.}~\bibnamefont {Forets}}, \
  and\ \bibinfo {author} {\bibfnamefont {V.}~\bibnamefont {Nesme}},\
  }\href@noop {} {\enquote {\bibinfo {title} {{The Dirac equation as a Quantum
  Walk: higher-dimensions, convergence}},}\ } (\bibinfo {year} {2013}),\
  \bibinfo {note} {pre-print arXiv:1307.3524}\BibitemShut {NoStop}%
\bibitem [{\citenamefont {Arrighi}\ \emph {et~al.}(2014)\citenamefont
  {Arrighi}, \citenamefont {Facchini},\ and\ \citenamefont
  {Forets}}]{arrighi2014discrete}%
  \BibitemOpen
  \bibfield  {author} {\bibinfo {author} {\bibfnamefont {P.}~\bibnamefont
  {Arrighi}}, \bibinfo {author} {\bibfnamefont {S.}~\bibnamefont {Facchini}}, \
  and\ \bibinfo {author} {\bibfnamefont {M.}~\bibnamefont {Forets}},\
  }\href@noop {} {\bibfield  {journal} {\bibinfo  {journal} {New Journal of
  Physics}\ }\textbf {\bibinfo {volume} {16}},\ \bibinfo {pages} {093007}
  (\bibinfo {year} {2014})}\BibitemShut {NoStop}%
\bibitem [{\citenamefont {Bisio}\ \emph {et~al.}(2017)\citenamefont {Bisio},
  \citenamefont {D~Ariano},\ and\ \citenamefont {Perinotti}}]{DArianoLorentz}%
  \BibitemOpen
  \bibfield  {author} {\bibinfo {author} {\bibfnamefont {A.}~\bibnamefont
  {Bisio}}, \bibinfo {author} {\bibfnamefont {G.~M.}\ \bibnamefont {D~Ariano}},
  \ and\ \bibinfo {author} {\bibfnamefont {P.}~\bibnamefont {Perinotti}},\
  }\href@noop {} {\bibfield  {journal} {\bibinfo  {journal} {Foundations of
  Physics}\ }\textbf {\bibinfo {volume} {47}},\ \bibinfo {pages} {1065}
  (\bibinfo {year} {2017})}\BibitemShut {NoStop}%
\bibitem [{\citenamefont {Lloyd}(2005)}]{LloydQG}%
  \BibitemOpen
  \bibfield  {author} {\bibinfo {author} {\bibfnamefont {S.}~\bibnamefont
  {Lloyd}},\ }\href@noop {} {\enquote {\bibinfo {title} {{A theory of quantum
  gravity based on quantum computation}},}\ }\bibinfo {howpublished} {ArXiv
  preprint: quant-ph/0501135} (\bibinfo {year} {2005})\BibitemShut {NoStop}%
\bibitem [{\citenamefont {Neto}\ \emph {et~al.}(2009)\citenamefont {Neto},
  \citenamefont {Guinea}, \citenamefont {Peres}, \citenamefont {Novoselov},\
  and\ \citenamefont {Geim}}]{neto2009electronic}%
  \BibitemOpen
  \bibfield  {author} {\bibinfo {author} {\bibfnamefont {A.~C.}\ \bibnamefont
  {Neto}}, \bibinfo {author} {\bibfnamefont {F.}~\bibnamefont {Guinea}},
  \bibinfo {author} {\bibfnamefont {N.~M.}\ \bibnamefont {Peres}}, \bibinfo
  {author} {\bibfnamefont {K.~S.}\ \bibnamefont {Novoselov}}, \ and\ \bibinfo
  {author} {\bibfnamefont {A.~K.}\ \bibnamefont {Geim}},\ }\href@noop {}
  {\bibfield  {journal} {\bibinfo  {journal} {Reviews of modern physics}\
  }\textbf {\bibinfo {volume} {81}},\ \bibinfo {pages} {109} (\bibinfo {year}
  {2009})}\BibitemShut {NoStop}%
\bibitem [{\citenamefont {Ye}\ \emph {et~al.}(2018)\citenamefont {Ye},
  \citenamefont {Kang}, \citenamefont {Liu}, \citenamefont {von Cube},
  \citenamefont {Wicker}, \citenamefont {Suzuki}, \citenamefont {Jozwiak},
  \citenamefont {Bostwick}, \citenamefont {Rotenberg}, \citenamefont {Bell},
  \citenamefont {Fu}, \citenamefont {Comin},\ and\ \citenamefont
  {Checkelsky}}]{Ye2018}%
  \BibitemOpen
  \bibfield  {author} {\bibinfo {author} {\bibfnamefont {L.}~\bibnamefont
  {Ye}}, \bibinfo {author} {\bibfnamefont {M.}~\bibnamefont {Kang}}, \bibinfo
  {author} {\bibfnamefont {J.}~\bibnamefont {Liu}}, \bibinfo {author}
  {\bibfnamefont {F.}~\bibnamefont {von Cube}}, \bibinfo {author}
  {\bibfnamefont {C.~R.}\ \bibnamefont {Wicker}}, \bibinfo {author}
  {\bibfnamefont {T.}~\bibnamefont {Suzuki}}, \bibinfo {author} {\bibfnamefont
  {C.}~\bibnamefont {Jozwiak}}, \bibinfo {author} {\bibfnamefont
  {A.}~\bibnamefont {Bostwick}}, \bibinfo {author} {\bibfnamefont
  {E.}~\bibnamefont {Rotenberg}}, \bibinfo {author} {\bibfnamefont {D.~C.}\
  \bibnamefont {Bell}}, \bibinfo {author} {\bibfnamefont {L.}~\bibnamefont
  {Fu}}, \bibinfo {author} {\bibfnamefont {R.}~\bibnamefont {Comin}}, \ and\
  \bibinfo {author} {\bibfnamefont {J.~G.}\ \bibnamefont {Checkelsky}},\ }\href
  {https://doi.org/10.1038/nature25987} {\bibfield  {journal} {\bibinfo
  {journal} {Nature}\ }\textbf {\bibinfo {volume} {555}},\ \bibinfo {pages}
  {638} (\bibinfo {year} {2018})}\BibitemShut {NoStop}%
\bibitem [{\citenamefont {Bougroura}\ \emph {et~al.}(2016)\citenamefont
  {Bougroura}, \citenamefont {Aissaoui}, \citenamefont {Chancellor},\ and\
  \citenamefont {Kendon}}]{Bougroura2016}%
  \BibitemOpen
  \bibfield  {author} {\bibinfo {author} {\bibfnamefont {H.}~\bibnamefont
  {Bougroura}}, \bibinfo {author} {\bibfnamefont {H.}~\bibnamefont {Aissaoui}},
  \bibinfo {author} {\bibfnamefont {N.}~\bibnamefont {Chancellor}}, \ and\
  \bibinfo {author} {\bibfnamefont {V.}~\bibnamefont {Kendon}},\ }\href
  {\doibase 10.1103/PhysRevA.94.062331} {\bibfield  {journal} {\bibinfo
  {journal} {Physical Review A}\ }\textbf {\bibinfo {volume} {94}},\ \bibinfo
  {pages} {1} (\bibinfo {year} {2016})},\ \Eprint
  {http://arxiv.org/abs/arXiv:1611.02991v1} {arXiv:arXiv:1611.02991v1}
  \BibitemShut {NoStop}%
\bibitem [{\citenamefont {Kitagawa}\ \emph {et~al.}(2010)\citenamefont
  {Kitagawa}, \citenamefont {Rudner}, \citenamefont {Berg},\ and\ \citenamefont
  {Demler}}]{Kitagawa2010}%
  \BibitemOpen
  \bibfield  {author} {\bibinfo {author} {\bibfnamefont {T.}~\bibnamefont
  {Kitagawa}}, \bibinfo {author} {\bibfnamefont {M.~S.}\ \bibnamefont
  {Rudner}}, \bibinfo {author} {\bibfnamefont {E.}~\bibnamefont {Berg}}, \ and\
  \bibinfo {author} {\bibfnamefont {E.}~\bibnamefont {Demler}},\ }\href
  {\doibase 10.1103/PhysRevA.82.033429} {\bibfield  {journal} {\bibinfo
  {journal} {Physical Review A - Atomic, Molecular, and Optical Physics}\
  }\textbf {\bibinfo {volume} {82}} (\bibinfo {year} {2010}),\
  10.1103/PhysRevA.82.033429},\ \Eprint {http://arxiv.org/abs/1003.1729}
  {arXiv:1003.1729} \BibitemShut {NoStop}%
\bibitem [{\citenamefont {Regge}(1961)}]{Regge1961}%
  \BibitemOpen
  \bibfield  {author} {\bibinfo {author} {\bibfnamefont {T.}~\bibnamefont
  {Regge}},\ }\href {\doibase 10.1007/BF02733251} {\bibfield  {journal}
  {\bibinfo  {journal} {Il Nuovo Cimento (1955-1965)}\ }\textbf {\bibinfo
  {volume} {19}},\ \bibinfo {pages} {558} (\bibinfo {year} {1961})}\BibitemShut
  {NoStop}%
\bibitem [{\citenamefont {Rovelli}(1998)}]{Rovelli1998}%
  \BibitemOpen
  \bibfield  {author} {\bibinfo {author} {\bibfnamefont {C.}~\bibnamefont
  {Rovelli}},\ }\href {\doibase 10.12942/lrr-1998-1} {\bibfield  {journal}
  {\bibinfo  {journal} {Living Reviews in Relativity}\ }\textbf {\bibinfo
  {volume} {1}},\ \bibinfo {pages} {1} (\bibinfo {year} {1998})}\BibitemShut
  {NoStop}%
\bibitem [{\citenamefont {Ambjorn}\ \emph {et~al.}(2006)\citenamefont
  {Ambjorn}, \citenamefont {Jurkiewicz},\ and\ \citenamefont
  {Loll}}]{Ambjorn2006}%
  \BibitemOpen
  \bibfield  {author} {\bibinfo {author} {\bibfnamefont {J.}~\bibnamefont
  {Ambjorn}}, \bibinfo {author} {\bibfnamefont {J.}~\bibnamefont {Jurkiewicz}},
  \ and\ \bibinfo {author} {\bibfnamefont {R.}~\bibnamefont {Loll}},\ }\href
  {\doibase 10.1080/00107510600603344} {\bibfield  {journal} {\bibinfo
  {journal} {Contemporary Physics}\ }\textbf {\bibinfo {volume} {47}},\
  \bibinfo {pages} {103} (\bibinfo {year} {2006})},\ \Eprint
  {http://arxiv.org/abs/https://doi.org/10.1080/00107510600603344}
  {https://doi.org/10.1080/00107510600603344} \BibitemShut {NoStop}%
\bibitem [{\citenamefont {Jay}\ \emph {et~al.}()\citenamefont {Jay},
  \citenamefont {Debbasch},\ and\ \citenamefont {Wang}}]{Jay2018}%
  \BibitemOpen
  \bibfield  {author} {\bibinfo {author} {\bibfnamefont {G.}~\bibnamefont
  {Jay}}, \bibinfo {author} {\bibfnamefont {F.}~\bibnamefont {Debbasch}}, \
  and\ \bibinfo {author} {\bibfnamefont {J.~B.}\ \bibnamefont {Wang}},\
  }\href@noop {} {\ }\Eprint {http://arxiv.org/abs/1803.01304v1} {1803.01304v1}
  \BibitemShut {NoStop}%
\bibitem [{\citenamefont {Succi}\ and\ \citenamefont
  {Benzi}(1993)}]{BenziSucci}%
  \BibitemOpen
  \bibfield  {author} {\bibinfo {author} {\bibfnamefont {S.}~\bibnamefont
  {Succi}}\ and\ \bibinfo {author} {\bibfnamefont {R.}~\bibnamefont {Benzi}},\
  }\href@noop {} {\bibfield  {journal} {\bibinfo  {journal} {Physica D:
  Nonlinear Phenomena}\ }\textbf {\bibinfo {volume} {69}},\ \bibinfo {pages}
  {327} (\bibinfo {year} {1993})}\BibitemShut {NoStop}%
\bibitem [{\citenamefont {Dellar}\ \emph {et~al.}(2011)\citenamefont {Dellar},
  \citenamefont {Lapitski}, \citenamefont {Palpacelli},\ and\ \citenamefont
  {Succi}}]{LapitskiDellarPalpacelliSucci}%
  \BibitemOpen
  \bibfield  {author} {\bibinfo {author} {\bibfnamefont {P.~J.}\ \bibnamefont
  {Dellar}}, \bibinfo {author} {\bibfnamefont {D.}~\bibnamefont {Lapitski}},
  \bibinfo {author} {\bibfnamefont {S.}~\bibnamefont {Palpacelli}}, \ and\
  \bibinfo {author} {\bibfnamefont {S.}~\bibnamefont {Succi}},\ }\href
  {\doibase 10.1103/PhysRevE.83.046706} {\bibfield  {journal} {\bibinfo
  {journal} {Phys. Rev. E}\ }\textbf {\bibinfo {volume} {83}},\ \bibinfo
  {pages} {046706} (\bibinfo {year} {2011})}\BibitemShut {NoStop}%
\bibitem [{\citenamefont {Bisio}\ \emph {et~al.}(2012)\citenamefont {Bisio},
  \citenamefont {D'Ariano},\ and\ \citenamefont {Tosini}}]{DAriano}%
  \BibitemOpen
  \bibfield  {author} {\bibinfo {author} {\bibfnamefont {A.}~\bibnamefont
  {Bisio}}, \bibinfo {author} {\bibfnamefont {G.~M.}\ \bibnamefont {D'Ariano}},
  \ and\ \bibinfo {author} {\bibfnamefont {A.}~\bibnamefont {Tosini}},\
  }\href@noop {} {\bibfield  {journal} {\bibinfo  {journal} {arXiv preprint
  arXiv:1212.2839}\ } (\bibinfo {year} {2012})}\BibitemShut {NoStop}%
\bibitem [{\citenamefont {Chandrashekar}(2013)}]{chandrashekar2013two}%
  \BibitemOpen
  \bibfield  {author} {\bibinfo {author} {\bibfnamefont {C.}~\bibnamefont
  {Chandrashekar}},\ }\href@noop {} {\bibfield  {journal} {\bibinfo  {journal}
  {Scientific reports}\ }\textbf {\bibinfo {volume} {3}},\ \bibinfo {pages}
  {2829} (\bibinfo {year} {2013})}\BibitemShut {NoStop}%
\bibitem [{\citenamefont {Chandrashekar}\ \emph {et~al.}(2010)\citenamefont
  {Chandrashekar}, \citenamefont {Banerjee},\ and\ \citenamefont
  {Srikanth}}]{IndiansDirac}%
  \BibitemOpen
  \bibfield  {author} {\bibinfo {author} {\bibfnamefont {C.}~\bibnamefont
  {Chandrashekar}}, \bibinfo {author} {\bibfnamefont {S.}~\bibnamefont
  {Banerjee}}, \ and\ \bibinfo {author} {\bibfnamefont {R.}~\bibnamefont
  {Srikanth}},\ }\href@noop {} {\bibfield  {journal} {\bibinfo  {journal}
  {Phys. Rev. A.}\ }\textbf {\bibinfo {volume} {81}},\ \bibinfo {pages} {62340}
  (\bibinfo {year} {2010})}\BibitemShut {NoStop}%
\bibitem [{\citenamefont {Arrighi}\ and\ \citenamefont
  {Facchini}(2013)}]{ArrighiKG}%
  \BibitemOpen
  \bibfield  {author} {\bibinfo {author} {\bibfnamefont {P.}~\bibnamefont
  {Arrighi}}\ and\ \bibinfo {author} {\bibfnamefont {S.}~\bibnamefont
  {Facchini}},\ }\href@noop {} {\bibfield  {journal} {\bibinfo  {journal} {EPL
  (Europhysics Letters)}\ }\textbf {\bibinfo {volume} {104}},\ \bibinfo {pages}
  {60004} (\bibinfo {year} {2013})}\BibitemShut {NoStop}%
\bibitem [{\citenamefont {di~Molfetta}\ and\ \citenamefont
  {Debbasch}(2012)}]{MolfettaDebbasch}%
  \BibitemOpen
  \bibfield  {author} {\bibinfo {author} {\bibfnamefont {G.}~\bibnamefont
  {di~Molfetta}}\ and\ \bibinfo {author} {\bibfnamefont {F.}~\bibnamefont
  {Debbasch}},\ }\href@noop {} {\bibfield  {journal} {\bibinfo  {journal}
  {Journal of Mathematical Physics}\ }\textbf {\bibinfo {volume} {53}},\
  \bibinfo {pages} {123302} (\bibinfo {year} {2012})}\BibitemShut {NoStop}%
\bibitem [{\citenamefont {Strauch}(2006)}]{StrauchShrodinger}%
  \BibitemOpen
  \bibfield  {author} {\bibinfo {author} {\bibfnamefont {F.~W.}\ \bibnamefont
  {Strauch}},\ }\href@noop {} {\bibfield  {journal} {\bibinfo  {journal}
  {Physical Review A}\ }\textbf {\bibinfo {volume} {73}},\ \bibinfo {pages}
  {054302} (\bibinfo {year} {2006})}\BibitemShut {NoStop}%
\bibitem [{\citenamefont {Love}\ and\ \citenamefont
  {Boghosian}(2005)}]{LoveBoghosian}%
  \BibitemOpen
  \bibfield  {author} {\bibinfo {author} {\bibfnamefont {P.}~\bibnamefont
  {Love}}\ and\ \bibinfo {author} {\bibfnamefont {B.}~\bibnamefont
  {Boghosian}},\ }\href@noop {} {\bibfield  {journal} {\bibinfo  {journal}
  {Quantum Information Processing}\ }\textbf {\bibinfo {volume} {4}},\ \bibinfo
  {pages} {335} (\bibinfo {year} {2005})}\BibitemShut {NoStop}%
\bibitem [{\citenamefont {Cedzich}\ \emph {et~al.}(2013)\citenamefont
  {Cedzich}, \citenamefont {Ryb{\'a}r}, \citenamefont {Werner}, \citenamefont
  {Alberti}, \citenamefont {Genske},\ and\ \citenamefont
  {Werner}}]{cedzich2013propagation}%
  \BibitemOpen
  \bibfield  {author} {\bibinfo {author} {\bibfnamefont {C.}~\bibnamefont
  {Cedzich}}, \bibinfo {author} {\bibfnamefont {T.}~\bibnamefont {Ryb{\'a}r}},
  \bibinfo {author} {\bibfnamefont {A.}~\bibnamefont {Werner}}, \bibinfo
  {author} {\bibfnamefont {A.}~\bibnamefont {Alberti}}, \bibinfo {author}
  {\bibfnamefont {M.}~\bibnamefont {Genske}}, \ and\ \bibinfo {author}
  {\bibfnamefont {R.}~\bibnamefont {Werner}},\ }\href@noop {} {\bibfield
  {journal} {\bibinfo  {journal} {Physical review letters}\ }\textbf {\bibinfo
  {volume} {111}},\ \bibinfo {pages} {160601} (\bibinfo {year}
  {2013})}\BibitemShut {NoStop}%
\bibitem [{\citenamefont {Di~Molfetta}\ \emph {et~al.}(2014)\citenamefont
  {Di~Molfetta}, \citenamefont {Brachet},\ and\ \citenamefont
  {Debbasch}}]{di2014quantum}%
  \BibitemOpen
  \bibfield  {author} {\bibinfo {author} {\bibfnamefont {G.}~\bibnamefont
  {Di~Molfetta}}, \bibinfo {author} {\bibfnamefont {M.}~\bibnamefont
  {Brachet}}, \ and\ \bibinfo {author} {\bibfnamefont {F.}~\bibnamefont
  {Debbasch}},\ }\href@noop {} {\bibfield  {journal} {\bibinfo  {journal}
  {Physica A: Statistical Mechanics and its Applications}\ }\textbf {\bibinfo
  {volume} {397}},\ \bibinfo {pages} {157} (\bibinfo {year}
  {2014})}\BibitemShut {NoStop}%
\bibitem [{\citenamefont {M{\'a}rquez-Mart{\'\i}n}\ \emph
  {et~al.}(2017)\citenamefont {M{\'a}rquez-Mart{\'\i}n}, \citenamefont
  {Di~Molfetta},\ and\ \citenamefont {P{\'e}rez}}]{marquez2017fermion}%
  \BibitemOpen
  \bibfield  {author} {\bibinfo {author} {\bibfnamefont {I.}~\bibnamefont
  {M{\'a}rquez-Mart{\'\i}n}}, \bibinfo {author} {\bibfnamefont
  {G.}~\bibnamefont {Di~Molfetta}}, \ and\ \bibinfo {author} {\bibfnamefont
  {A.}~\bibnamefont {P{\'e}rez}},\ }\href {\doibase
  https://doi.org/10.1103/PhysRevA.95.042112} {\bibfield  {journal} {\bibinfo
  {journal} {Physical Review A}\ }\textbf {\bibinfo {volume} {95}},\ \bibinfo
  {pages} {042112} (\bibinfo {year} {2017})}\BibitemShut {NoStop}%
\bibitem [{\citenamefont {Di~Molfetta}\ and\ \citenamefont
  {P{\'e}rez}(2016)}]{di2016quantum}%
  \BibitemOpen
  \bibfield  {author} {\bibinfo {author} {\bibfnamefont {G.}~\bibnamefont
  {Di~Molfetta}}\ and\ \bibinfo {author} {\bibfnamefont {A.}~\bibnamefont
  {P{\'e}rez}},\ }\href {\doibase
  https://doi.org/10.1088/1367-2630/18/10/103038} {\bibfield  {journal}
  {\bibinfo  {journal} {New Journal of Physics}\ }\textbf {\bibinfo {volume}
  {18}},\ \bibinfo {pages} {103038} (\bibinfo {year} {2016})}\BibitemShut
  {NoStop}%
\bibitem [{\citenamefont {Arnault}\ \emph {et~al.}(2016)\citenamefont
  {Arnault}, \citenamefont {Di~Molfetta}, \citenamefont {Brachet},\ and\
  \citenamefont {Debbasch}}]{arnault2016quantum}%
  \BibitemOpen
  \bibfield  {author} {\bibinfo {author} {\bibfnamefont {P.}~\bibnamefont
  {Arnault}}, \bibinfo {author} {\bibfnamefont {G.}~\bibnamefont
  {Di~Molfetta}}, \bibinfo {author} {\bibfnamefont {M.}~\bibnamefont
  {Brachet}}, \ and\ \bibinfo {author} {\bibfnamefont {F.}~\bibnamefont
  {Debbasch}},\ }\href {\doibase https://doi.org/10.1103/PhysRevA.94.012335}
  {\bibfield  {journal} {\bibinfo  {journal} {Physical Review A}\ }\textbf
  {\bibinfo {volume} {94}},\ \bibinfo {pages} {012335} (\bibinfo {year}
  {2016})}\BibitemShut {NoStop}%
\bibitem [{\citenamefont {Meyer}(1997)}]{meyer1997quantum}%
  \BibitemOpen
  \bibfield  {author} {\bibinfo {author} {\bibfnamefont {D.~A.}\ \bibnamefont
  {Meyer}},\ }\href@noop {} {\bibfield  {journal} {\bibinfo  {journal}
  {International Journal of Modern Physics C}\ }\textbf {\bibinfo {volume}
  {8}},\ \bibinfo {pages} {717} (\bibinfo {year} {1997})}\BibitemShut {NoStop}%
\bibitem [{\citenamefont {Ahlbrecht}\ \emph {et~al.}(2012)\citenamefont
  {Ahlbrecht}, \citenamefont {Alberti}, \citenamefont {Meschede}, \citenamefont
  {Scholz}, \citenamefont {Werner},\ and\ \citenamefont
  {Werner}}]{ahlbrecht2012molecular}%
  \BibitemOpen
  \bibfield  {author} {\bibinfo {author} {\bibfnamefont {A.}~\bibnamefont
  {Ahlbrecht}}, \bibinfo {author} {\bibfnamefont {A.}~\bibnamefont {Alberti}},
  \bibinfo {author} {\bibfnamefont {D.}~\bibnamefont {Meschede}}, \bibinfo
  {author} {\bibfnamefont {V.~B.}\ \bibnamefont {Scholz}}, \bibinfo {author}
  {\bibfnamefont {A.~H.}\ \bibnamefont {Werner}}, \ and\ \bibinfo {author}
  {\bibfnamefont {R.~F.}\ \bibnamefont {Werner}},\ }\href@noop {} {\bibfield
  {journal} {\bibinfo  {journal} {New Journal of Physics}\ }\textbf {\bibinfo
  {volume} {14}},\ \bibinfo {pages} {073050} (\bibinfo {year}
  {2012})}\BibitemShut {NoStop}%
\bibitem [{\citenamefont {Di~Molfetta}\ \emph {et~al.}(2013)\citenamefont
  {Di~Molfetta}, \citenamefont {Brachet},\ and\ \citenamefont
  {Debbasch}}]{di2013quantum}%
  \BibitemOpen
  \bibfield  {author} {\bibinfo {author} {\bibfnamefont {G.}~\bibnamefont
  {Di~Molfetta}}, \bibinfo {author} {\bibfnamefont {M.}~\bibnamefont
  {Brachet}}, \ and\ \bibinfo {author} {\bibfnamefont {F.}~\bibnamefont
  {Debbasch}},\ }\href@noop {} {\bibfield  {journal} {\bibinfo  {journal}
  {Physical Review A}\ }\textbf {\bibinfo {volume} {88}},\ \bibinfo {pages}
  {042301} (\bibinfo {year} {2013})}\BibitemShut {NoStop}%
\bibitem [{\citenamefont {Arrighi}\ \emph {et~al.}(2016)\citenamefont
  {Arrighi}, \citenamefont {Facchini},\ and\ \citenamefont
  {Forets}}]{ArrighiGRDirac}%
  \BibitemOpen
  \bibfield  {author} {\bibinfo {author} {\bibfnamefont {P.}~\bibnamefont
  {Arrighi}}, \bibinfo {author} {\bibfnamefont {S.}~\bibnamefont {Facchini}}, \
  and\ \bibinfo {author} {\bibfnamefont {M.}~\bibnamefont {Forets}},\
  }\href@noop {} {\bibfield  {journal} {\bibinfo  {journal} {Quantum
  Information Processing}\ }\textbf {\bibinfo {volume} {15}},\ \bibinfo {pages}
  {3467} (\bibinfo {year} {2016})}\BibitemShut {NoStop}%
\bibitem [{\citenamefont {Arnault}\ and\ \citenamefont
  {Debbasch}(2017)}]{DebbaschWaves}%
  \BibitemOpen
  \bibfield  {author} {\bibinfo {author} {\bibfnamefont {P.}~\bibnamefont
  {Arnault}}\ and\ \bibinfo {author} {\bibfnamefont {F.}~\bibnamefont
  {Debbasch}},\ }\href {\doibase http://dx.doi.org/10.1016/j.aop.2017.04.003}
  {\bibfield  {journal} {\bibinfo  {journal} {Annals of Physics}\ }\textbf
  {\bibinfo {volume} {383}},\ \bibinfo {pages} {645 } (\bibinfo {year}
  {2017})}\BibitemShut {NoStop}%
\bibitem [{\citenamefont {Arrighi}\ and\ \citenamefont
  {Facchini}(2017)}]{ArrighiGRDirac3D}%
  \BibitemOpen
  \bibfield  {author} {\bibinfo {author} {\bibfnamefont {P.}~\bibnamefont
  {Arrighi}}\ and\ \bibinfo {author} {\bibfnamefont {F.}~\bibnamefont
  {Facchini}},\ }\href {https://arxiv.org/abs/1609.00305} {\bibfield  {journal}
  {\bibinfo  {journal} {Quantum Information and Computation}\ }\textbf
  {\bibinfo {volume} {17}},\ \bibinfo {pages} {0810} (\bibinfo {year}
  {2017})},\ \bibinfo {note} {arXiv:1609.00305}\BibitemShut {NoStop}%
\bibitem [{\citenamefont {Stegmann}\ and\ \citenamefont
  {Szpak}(2016)}]{stegmann2016current}%
  \BibitemOpen
  \bibfield  {author} {\bibinfo {author} {\bibfnamefont {T.}~\bibnamefont
  {Stegmann}}\ and\ \bibinfo {author} {\bibfnamefont {N.}~\bibnamefont
  {Szpak}},\ }\href@noop {} {\bibfield  {journal} {\bibinfo  {journal} {New
  Journal of Physics}\ }\textbf {\bibinfo {volume} {18}},\ \bibinfo {pages}
  {053016} (\bibinfo {year} {2016})}\BibitemShut {NoStop}%
\bibitem [{\citenamefont {Kerner}\ \emph {et~al.}(2012)\citenamefont {Kerner},
  \citenamefont {Naumis},\ and\ \citenamefont
  {G{\'o}mez-Arias}}]{kerner2012bending}%
  \BibitemOpen
  \bibfield  {author} {\bibinfo {author} {\bibfnamefont {R.}~\bibnamefont
  {Kerner}}, \bibinfo {author} {\bibfnamefont {G.~G.}\ \bibnamefont {Naumis}},
  \ and\ \bibinfo {author} {\bibfnamefont {W.~A.}\ \bibnamefont
  {G{\'o}mez-Arias}},\ }\href@noop {} {\bibfield  {journal} {\bibinfo
  {journal} {Physica B: Condensed Matter}\ }\textbf {\bibinfo {volume} {407}},\
  \bibinfo {pages} {2002} (\bibinfo {year} {2012})}\BibitemShut {NoStop}%
\bibitem [{\citenamefont {Abal}\ \emph {et~al.}(2010)\citenamefont {Abal},
  \citenamefont {Donangelo}, \citenamefont {Marquezino},\ and\ \citenamefont
  {Portugal}}]{Abal2010}%
  \BibitemOpen
  \bibfield  {author} {\bibinfo {author} {\bibfnamefont {G.}~\bibnamefont
  {Abal}}, \bibinfo {author} {\bibfnamefont {R.}~\bibnamefont {Donangelo}},
  \bibinfo {author} {\bibfnamefont {F.~L.}\ \bibnamefont {Marquezino}}, \ and\
  \bibinfo {author} {\bibfnamefont {R.}~\bibnamefont {Portugal}},\ }\href
  {\doibase 10.1017/S0960129510000332} {\bibfield  {journal} {\bibinfo
  {journal} {Mathematical Structures in Computer Science}\ }\textbf {\bibinfo
  {volume} {20}},\ \bibinfo {pages} {999} (\bibinfo {year} {2010})},\ \Eprint
  {http://arxiv.org/abs/1001.1139} {arXiv:1001.1139} \BibitemShut {NoStop}%
\bibitem [{\citenamefont {Foulger}\ \emph {et~al.}(2015)\citenamefont
  {Foulger}, \citenamefont {Gnutzmann},\ and\ \citenamefont
  {Tanner}}]{Foulger2015}%
  \BibitemOpen
  \bibfield  {author} {\bibinfo {author} {\bibfnamefont {I.}~\bibnamefont
  {Foulger}}, \bibinfo {author} {\bibfnamefont {S.}~\bibnamefont {Gnutzmann}},
  \ and\ \bibinfo {author} {\bibfnamefont {G.}~\bibnamefont {Tanner}},\ }\href
  {\doibase 10.1103/PhysRevA.91.062323} {\bibfield  {journal} {\bibinfo
  {journal} {Physical Review A - Atomic, Molecular, and Optical Physics}\
  }\textbf {\bibinfo {volume} {91}},\ \bibinfo {pages} {1} (\bibinfo {year}
  {2015})},\ \Eprint {http://arxiv.org/abs/arXiv:1312.3852v1}
  {arXiv:arXiv:1312.3852v1} \BibitemShut {NoStop}%
\bibitem [{\citenamefont {Karafyllidis}(2015)}]{Karafyllidis2015}%
  \BibitemOpen
  \bibfield  {author} {\bibinfo {author} {\bibfnamefont {I.~G.}\ \bibnamefont
  {Karafyllidis}},\ }\href@noop {} {\bibfield  {journal} {\bibinfo  {journal}
  {Journal of Computational Science}\ }\textbf {\bibinfo {volume} {11}},\
  \bibinfo {pages} {326} (\bibinfo {year} {2015})}\BibitemShut {NoStop}%
\bibitem [{\citenamefont {Lawrie}(2001)}]{Lawrie}%
  \BibitemOpen
  \bibfield  {author} {\bibinfo {author} {\bibfnamefont {I.~D.}\ \bibnamefont
  {Lawrie}},\ }\href {https://www.taylorfrancis.com/books/9781439858196} {\emph
  {\bibinfo {title} {Unified grand tour of theoretical physics}}},\ \bibinfo
  {edition} {2nd}\ ed.\ (\bibinfo  {publisher} {Taylor \& Francis, New York},\
  \bibinfo {year} {2001})\BibitemShut {NoStop}%
\bibitem [{\citenamefont {Koke}\ \emph {et~al.}(2016)\citenamefont {Koke},
  \citenamefont {Noh},\ and\ \citenamefont {Angelakis}}]{Koke2016}%
  \BibitemOpen
  \bibfield  {author} {\bibinfo {author} {\bibfnamefont {C.}~\bibnamefont
  {Koke}}, \bibinfo {author} {\bibfnamefont {C.}~\bibnamefont {Noh}}, \ and\
  \bibinfo {author} {\bibfnamefont {D.~G.}\ \bibnamefont {Angelakis}},\ }\href
  {\doibase https://doi.org/10.1016/j.aop.2016.08.013} {\bibfield  {journal}
  {\bibinfo  {journal} {Annals of Physics}\ }\textbf {\bibinfo {volume}
  {374}},\ \bibinfo {pages} {162 } (\bibinfo {year} {2016})}\BibitemShut
  {NoStop}%
\bibitem [{\citenamefont {Yepez}()}]{Yepez2011}%
  \BibitemOpen
  \bibfield  {author} {\bibinfo {author} {\bibfnamefont {J.}~\bibnamefont
  {Yepez}},\ }\href@noop {} {\ }\Eprint {http://arxiv.org/abs/1106.2037v1}
  {1106.2037v1} \BibitemShut {NoStop}%
\bibitem [{\citenamefont {De~Oliveira}\ and\ \citenamefont
  {Tiomno}(1962)}]{de1962representations}%
  \BibitemOpen
  \bibfield  {author} {\bibinfo {author} {\bibfnamefont {C.}~\bibnamefont
  {De~Oliveira}}\ and\ \bibinfo {author} {\bibfnamefont {J.}~\bibnamefont
  {Tiomno}},\ }\href@noop {} {\bibfield  {journal} {\bibinfo  {journal} {Il
  Nuovo Cimento}\ }\textbf {\bibinfo {volume} {24}},\ \bibinfo {pages} {672}
  (\bibinfo {year} {1962})}\BibitemShut {NoStop}%
\end{thebibliography}
\end{document}